➔ *Regular Research Paper – NS*

# mHealth: An Artificial Intelligence Oriented Mobile Application for Personal Healthcare Support

**Ismail Ali AFRAH**
Suleyman Demirel University, Turkey
haajihagaa@gmail.com

**Utku KOSE**
Suleyman Demirel University, Turkey
utkukose@sdu.edu.tr

## Abstract

Main objective of this study is to introduce an expert system-based mHealth application that takes Artificial Intelligence support by considering previously introduced solutions from the literature and employing possible requirements for a better solution. Thanks to that research study, a mobile software system having Artificial Intelligence support and providing dynamic support against the common health problems in daily life was designed-developed and it was evaluated via survey and diagnosis-based evaluation tasks. Evaluation tasks indicated positive outcomes for the mHealth system.

**[This article was produced from the MSc. thesis (Author: Ismail Ali AFRAH) entitled as "DEVELOPING AN ARTIFICIAL INTELLIGENCE BASED MOBILE EXPERT SYSTEM APPLICATION FOR PERSONAL HEALTH SUPPORT" as accepted in the Dept. of Computer Engineering of Institute of Natural Sciences, Suleyman Demirel University, Turkey.]**

## 1. INTRODUCTION

Technology is an important tool for human life to keep up with the changes in the world and the universe. When viewed abstractly, all kinds of knowledge and skill accumulations that humanity has acquired in the direction of discoveries and inventions can be accepted within the scope of technology, and in concrete terms, all kinds of tools and equipment that will provide the pace in the best way are known as technologies. Technological developments, contrary to previous periods, have reached a very high acceleration since the beginning of the 20th century. With the advancement of technology, people's understanding of the world and the universe has ultimately required them to develop effective technological solutions by understanding each other. In addition, the power and capacity of human beings to produce solutions within their limits have led to the development of technologies that will support these skills. The most prominent technologies in question are known as computer and communication technologies [1-3].

The development of computer and communication technologies has led to the introduction of various technological tools and applications that will support people both in daily life and in general terms. These tools and practices have enabled humanity to move forward in different fields and even to reveal new areas depending on the progress. When viewed in a sectoral sense, computer and communication technologies have become associated with almost every sector today. The practices developed and put forward also manifest themselves in the health sector. The health sector, which is of great importance in terms of human life, can transfer the latest developments to real world applications in a short time as a sector that is primarily affected by technological developments. One of the developing application types in the health sector is mobile-based personal health support assistants. The number of mobile personal health support assistants, whose name is mHealth in the literature, is increasing day by day [4, 5].

The need for an overview of what has been revealed about mobile personal health support assistants in the healthcare industry is critical to developing alternative applications. At this point, it is beneficial to reveal the tendency towards developing an alternative application by using known applications. Accordingly, mHealth applications are software that can be downloaded from different online stores to smart devices such as tablets, smartphones and smart watches. There are also health applications on websites adapted for mobile devices. However, mobile health and health applications that can be downloaded directly to a mobile device are also considered as a part of mHealth concept. While some of the mobile health applications are simple tasks such as saving and storing data, some are applications that speed up the Artificial Intelligence-based process and make emergency calls [6-8]. For protecting people's health through smart phones or tablets, mHealth applications provide great convenience to individuals in terms of money.

Beyond all the mentioned technological features, mHealth applications always need renewal and original contributions under developing and changing conditions. Especially the efficient and productive use of more than one technology and scientific techniques, which have been proven to be successful today, in the mobile application environment, as much as possible, will provide highly qualified contributions to the literature. In detail, as there are many examples of Artificial Intelligence based solutions for medical-healthcare [9-11] and web based mobile applications or modular applications are in common place generally [12-14], it is a necessary to build a mobile solution accordingly. Because the Artificial Intelligence is the science of the future [15, 16], that's also a need for the humankind.

Based on the explanations, the aim of this study is to develop a mobile personal health application supported by the Expert System by making use of today's current technologies. The application to be developed in line with this purpose is aimed to have a structure that operates through central management, providing cloud server support and a secure communication infrastructure in a divisional system approach. Again, the technical infrastructure that will provide personal health support is aimed to use Artificial Neural Networks (ANN) and Intelligent Optimization techniques, as well as Image Processing, Signal Processing and Data Mining mechanisms, together with the Expert System technique. The results to be achieved with the study are briefly as follows:

• It is thought that the study will provide a comprehensive, effective and efficient mobile level personal health assistant application, by employing advanced components such as Artificial Intelligence, Image Processing, and the relevant technologies.

• It will cause establishing a practice that will improve - improve personal health awareness, which is often neglected.

• It is thought that the outputs will be obtained within the scope of widespread effective:

• The mHealth application is thought to be a great gain for the health sector and personal needs in the context of healthcare.

• Although the mentioned mHealth-focused application is a unique output in terms of its usage features and functions, it will enable the development of the associated literature.

• The application to be developed will be shared with relevant health institutions at national and international level, and thus it will be put into use and spread as quickly as possible.

• The performed study and the mHealth application are both thought to be increased personal health awareness and the effectiveness of the relevant health institutions in this context.

## 2.    DESIGN AND DEVELOPMENT PROCESS

There is a mobile application development process in this study. Accordingly, the general development steps of the application and the sub-studies that form the basis of the study include the following general phases that follow each other:

• **Step 1 / Needs analysis:** Analyzing and evaluating the needs for personal health and mobile application use during the development phase of the application.

• **Step 2 / Algorithmic infrastructure design - modeling:** The design and modeling of relevant algorithms, taking into account the organizational structure of the application, and making necessary updates according to their complexity.

• **Step 3 / Design - modeling of supporting technologies:** It will be involved in implementation; Designing, modeling and updating the related algorithms and techniques focused on Artificial Intelligence, Image / Signal Processing, Intelligent Optimization and Data Mining according to the evaluations.

• **Step 4 / Establishing and developing data infrastructure:** Designing the general database and data infrastructure system to be used by the application.

• **Step 5 / Expert information gathering processes:** It will be used for the solution of personal health-oriented problems; Obtaining the required expert knowledge from relevant authorities, compiling and applying them to practice.

• **Step 6 / Software design and coding processes:** The general outline of the application design and coding processes.

• **Step 7 / Test processes on the developed version:** Testing the obtained application pre-version.

• **Step 8 / Improvements - updates:** Improvements and updates needed within the application's testing processes.

• **Step 9 / Obtaining the latest application version:** Presenting the latest application version ready to use. Final tests on this version.

• **Step 10 / General evaluation and analysis in the direction of study:** Conducting general evaluation studies towards the scientific implications of the study, obtaining findings - results.

Within the scope of the mentioned stages, it is primarily aimed to introduce an Android-based mHealth application. For the purpose of the study, the Android version has been determined as the main priority since the applications in this operating system can be used easily and quickly.

### 2.1. Software Infrastructure and Supporting Technologies

When we examine the related literature, most of the applications put forward in mHealth and Artificial Intelligence scale are designed and developed around the applications considered in the previous section. In this study, based on both the studies described and similar applications in the literature, a suggestion for an Artificial Intelligence-based mobile personal health assistant application has been proposed. The details of the application are as follows:

#### 2.1.1. Division of labor system approach

The mobile application is built on a division of labor system framework, due to its Artificial Intelligence-based elements and its relationship with health data. Accordingly, the sections in the system can be explained as follows:

• **User Section:** The interface of the application that opens to the user - its hardware components are under this section.

• **Data Processing Module:** If the data coming from the user or the data to be presented to the user will be processed, it can pass through this section. There are mathematical algorithms, image signal processing algorithms and Data Mining algorithms that can be used for data pre-processing.

• **Communication Module:** If the application needs to communicate with the cloud environment, the necessary processes are carried out with this section.

• **Artificial Intelligence Module:** Artificial Intelligence infrastructure of the application will be located here. It will include PSO and CUR (optimization) along with ANN (Machine Learning).

• **Data - Expert Information Module:** User-related health data, application data and health-oriented expert information to be used when appropriate are under this section.

• **Management Department:** All components of the application that will be able to intervene at the management level (For example, updating - adding expert information, parameters of Artificial Intelligence techniques… etc.) Will be operated by the mechanisms under this section.

### 2.1.2. Central management and cloud server

In order for the mHealth application to be developed to work quickly and effectively, the organization of the described departments will be carried out by a system in a central position. This system will briefly be a collection of algorithms that direct the entire application process. Again, in addition to the central management approach, a mechanism for interaction with the cloud server will be added in order for the application to provide effective personal health assistance. The aim here is to keep the data on a remote server better and more secure, which may be insufficient for keeping and processing the application on the mobile device, and even apply to the cloud for more complex operations when necessary. In addition, this cloud server approach will help to keep increasing levels of personal data about users using the application. Although the efficiency and prevalence of cloud-oriented systems are on the surface, just like other studies in the literature, the advantages of this technology will be used in the application to be developed.

### 2.1.3. Secure communication

Communication processes in the mentioned mHealth application can be encrypted. The purpose here is to evaluate personal health data as private personal data, and therefore, to enable the encrypted exchange of data, especially in interactions with the cloud server. Likewise, users will be able to receive the feedback and assistance of smart personal health assistant supported by expert knowledge, with the option of personal - private use, thanks to various security-oriented approaches in the application process.

### 2.1.4. Supporting technologies and technical infrastructure

In order to provide an effective personal health system in a mobile environment, it was preferred to use a comprehensive technical infrastructure in practice. The technical infrastructure in question can be offered both embedded in the application and with cloud support within the possibilities of mobile devices. The details of the technical elements that will support Artificial Intelligence, connected techniques and other important functions of the application are briefly as follows:

• **Expert System:** The expert system [17] side of the application will be developed by adhering to the basics expressed in the literature. However, differently, a system in which the said expert system structure will be supported by alternative Artificial Intelligence techniques, Data Mining techniques and traditional data processing techniques will be presented. The expert system structure will be open to regulation through the Management Department of the application when necessary. These regulations will be supported with special data (eg specific disease symptoms, drug information… etc) to be added to the database infrastructure in addition to the classical "If… if… if not…" regulations.

• **Artificial Neural Networks (ANNs):** Thanks to Artificial Neural Networks [18], it has been ensured that various health data and personal information of the person using the application are blended and evaluated, and the interaction of the system with the person / study intensity is organized with criteria such as health score, diagnosis score, suggestion score. In addition, the Expert System infrastructure of the system has been configured to train and develop itself over ANN.

• **Intelligent Optimization:** In practice, intelligent optimization algorithms [19] were used to make decision mechanisms or some sensitive evaluations. Thus, optimization algorithms have produced rapid solutions in some application organizations and even health-oriented processes (analysis, diagnosis, etc.), where ANN or other algorithms will remain on a large scale and can only be solved by mathematical modeling. It is also important here whether optimization modeling is continuous optimization or combinatorial optimization. Since it is planned to benefit from both optimizations in the system, PSO and CUR, which are easy to code, were preferred.

• **Image Processing:** Processes focused on image processing [20] can take place as much as possible in the mobile device environment and analysis will be made possible through images that can be presented by the person and health interpretations can be made.

• **Signal Processing:** Signal processing [21] processes will also take place in the application environment in order to evaluate important signal-based data such as heart sound.

• **Data Mining Algorithms:** When appropriate, various Data Mining algorithms [22] will be included in the system environment to better process complex data by classifying or clustering.

## 2.2. Personal Health Assistant Approaches

In the developed application, personal health assistance processes targeted the following scale scope:

• Keeping various health-oriented characteristics of the person and providing feedback accordingly (For example, there will be a difference between a person with diabetes and a person without diabetes. Similar situation is valid for people with various allergic sensitivities).

• Making various constructive diagnoses by processing the data that the mobile device can receive based on sound, image or sensor.

• Providing feedback within the scope of various recommendations and expert knowledge in order to keep the person at a good level of health.

• Keeping records of health appointments and controls.

• Functions such as record keeping and reminders on processes such as taking medication and routine health checks.

• Advanced expert knowledge supported evaluations that can be made by interacting with the cloud server when necessary.

• Providing reports, statistical evaluations and opinions that can only come out with detailed analysis regarding the personal health status when necessary.

### 2.2.1. Daily health problems within the system

It is thought that the mHealth system to be developed will focus especially on daily health problems. At this point, besides the important health problems that concern people, there are diseases that have a very important place in daily life and reduce the quality of people's daily life routines or cause many advanced diseases when not intervened. In this context, the health problems-diseases within the scope of the study and the general symptoms considered are given in Table 1.

Table 1. Daily health problems in the mHealth system.

| Disease | Symptom |
|---|---|
| **Cold** | • High (for example, over 38ºC) fever<br>• Muscle pains<br>• Sweating<br>• Headache<br>• Fatigue / weakness<br>• Nasal congestion<br>• Throat ache |
| **Headaches** | • Pain in different parts of the head<br>• Neck tightness<br>• Shoulder stiffness<br>• Headache of varying duration and intensity |
| **Angina** | • Sensation of fluid flowing from the back of the throat (postnasal drip)<br>• Frequent throat clearing and sore throat<br>• Hoarseness |
| **Flu** | • A runny or stuffy nose<br>• Sensation of fluid flowing from the back of the throat (postnasal drip)<br>• Frequent throat clearing and sore throat<br>• Hoarseness<br>• Wheezing and shortness of breath |
| **Diabetes** | • Eating more than usual<br>• Weight gain<br>• Weakness and fatigue<br>• A runny or stuffy nose<br>• frequent urination<br>• Numbness in feet<br>• Fast and involuntary weight loss<br>• Blurred vision |
| **Viral Diseases** | • High fever<br>• Diarrhea<br>• Need for plenty of fluid<br>• High level of fatigue<br>• Wheezing and shortness of breath |
| **Vitamin Deficiency** | • Vision problems<br>• Feeling tired<br>• Mental fatigue<br>• Anemia<br>• Cracks in the skin<br>• Lifeless nails / hair<br>• Bone / muscle pains |

Also, some diseases that are integrated into mHealth and considered within the hormonal context and can be predicted with some basic symptoms in daily life are also expressed in Table 2 [23].

Table 2. Hormonal symptoms and diseases in mHealth system [23].

1- Weight gain and inability to lose weight: Insulin Resistance, Diabetes, Polycystic Ovary Syndrome, Cushing Syndrome and Hypothyroidism.

2- Sweet crises, frequent hunger: It can be the harbinger of hypoglycemia and insulin resistance.

3- Menstrual irregularity, increased hair growth: Ovarian cysts can be seen in adrenal gland disorders.

4- Purple striae (lines) in the abdomen: It may be a sign of Cushing Syndrome.

| |
|---|
| 5- <u>Fatigue:</u> It is one of the common symptoms of all hormonal diseases. |
| 6- <u>Swelling and pain in the neck:</u> Goiter may be a harbinger of thyroid nodule. |
| 7- <u>Palpitations, tremors in the hands:</u> It may be a symptom of hypoglycemia and overwork of the thyroid gland. |
| 8- <u>Excessive sweating:</u> Overwork of the thyroid may be a symptom of adrenal gland diseases and sugar drop. |
| 9- <u>Weight loss:</u> Diabetes can be a sign of overworking the thyroid. |
| 10- Body pain: It may be a symptom of parathyroid gland and vitamin D deficiency. |
| 11- <u>Bone melting, fracture in bones:</u> It may be a sign of osteoporosis. |
| 12- <u>Jaw enlargement:</u> May be a symptom of acromegaly. |
| 13- <u>Milk coming from the breasts:</u> It may be a symptom of prolactinoma. |
| 14- <u>Growth and Development Retardation:</u> Hormone deficiencies may be due to vitamin and iron deficiency. |
| 15- <u>No beard growth in men:</u> It may develop due to male hormone deficiency. |
| 16- <u>Impotence:</u> Testosterone hormone deficiency, excess prolactin, problems with the thyroid gland. |
| 17- <u>Breast growth:</u> It is called gynecomastia in men and should be investigated. |
| 18- <u>Excessive growth:</u> May be a sign of excess growth hormone. |
| 19- <u>Hair loss:</u> Polycystic ovary syndrome in women, adrenal gland disorders and thyroid gland problems can be seen. |
| 20- <u>Anemia:</u> It can be seen in hormone and vitamin deficiencies, problems related to thyroid and growth hormone. |
| 21- <u>Contraction and numbness in the hands:</u> Parathyroid gland, magnesium and calcium metabolism disorders and vitamin D deficiency can be seen. |
| 22- <u>High blood pressure:</u> Cushing's syndrome can be seen in parathyroid and thyroid gland disorders, adrenal gland disorders. |
| 23- <u>Skin thickening:</u> It may be a symptom of hypothyroidism. |
| 24- <u>Muscle twitching and pain:</u> Thyroid disorders can be seen in magnesium-calcium metabolism disorders. |

## 3. mHEALTH: ARTIFICIAL INTELLIGENCE BASED MOBILE EXPERT SYSTEM FOR HEALTH

Under this section, details about the mHealth system developed within the scope of the study are discussed. Accordingly, first of all, it was explained how the technologies - techniques and solution approaches discussed under the previous section were made functional in the relevant mobile application environment, and thus detailed information was given on the organization of the infrastructure of the system.

### 3.1. mHealth System General Operation Mechanisms

The developed mHealth system is generally based on the interaction of the user and the application via the mobile device, and it is also established on a system where cloud-sided data storage and data processing processes are operated. Accordingly, the mHealth system has been organized to operate on the following basic operation scenario in order to provide the user with an easy and fast usage experience:

• **If there is user-side triggering, follow the scenarios below:**

o Activating the relevant interfaces in case of using user-side application options.

o Activating the system's supportive technologies if there is a user-sided question - diagnosis process.

o If the use of application options is triggered, follow the scenarios below:

- ✓ Use of application control options.
- ✓ Use of application supporting technology options.

o If system supporting technologies are triggered, follow these scenarios:

- ✓ Perform the necessary algorithmic branching according to whether the received data is text (raw data), image or sound.
- ✓ Run data pre-processing modules if data pre-processing (normalization, transformation, etc.) after branching is required.
- ✓ If data pre-processing is required, such as image or signal processing, run the relevant modules.
- ✓ Determine which of the necessary Artificial Intelligence or Data Mining techniques to run if diagnosis - prediction processes are required after pre-processing.
- ✓ If the processes require intelligent optimization, determine the optimization mathematical models and determine which of the relevant techniques will be run for the solution.
- ✓ Expert knowledge in the process of running the relevant techniques, etc. If necessary, continue the communication process with the connected databases.

• **If there is no user-side triggering, follow the scenarios below:**

o Keep user personal health data up-to-date.

o Determine whether to give health-oriented feedback - alerts within the default control periods (or user-determined periods).

o Routine warnings, planned announcements - if there are warnings, execute.

o Continue logging (running errors, information, etc.) in the system background.

In the mHealth system, especially the diagnosis - estimation processes are considered as a process that requires careful processing of user-side data. For this reason, the system infrastructure enables the user to receive additional information or feedback depending on whether the data obtained from the external environment is related to the health problems kept on the database side, with or without user triggering. At this point, the process described below is operated directly:

*• Is the data of the retained user data set sufficient to diagnose any health problem?*

o If it is, go to the diagnosis - estimation processes.

o If not, request additional data from the user; otherwise, instruct the user to see a health facility.

### 3.2. mHealth Interfaces and General Usage

The interface structure of the developed mHealth system has been improved in order to provide an easy use. At this point, users who use the mHealth system for the first time must be registered in the system from two different profiles: "healthcare user" or "standard user". mHealth system can offer different types of interfaces and general usage patterns to two different types of users. Registration to the system is carried out through a combination of e-mail and password, and after the registration process, relevant database records and components of the user are set up in the AWS cloud system where the system is in communication. After users register to the system, they can login for use. After logging into the system, the general usage processes described under the following sub-headings can be performed:

### 3.2.1. Healthcare user interfaces and usage

The use of mHealth system by healthcare staff actually involves some kind of managerial processes. In this context, by using the menu displayed after logging in to the system, processes including system management, displaying information for users in the system, and even performing diagnostic processes that are more within the scope of standard users can be performed. Under Figure 1, mHealth system healthcare user menu screen is shown.

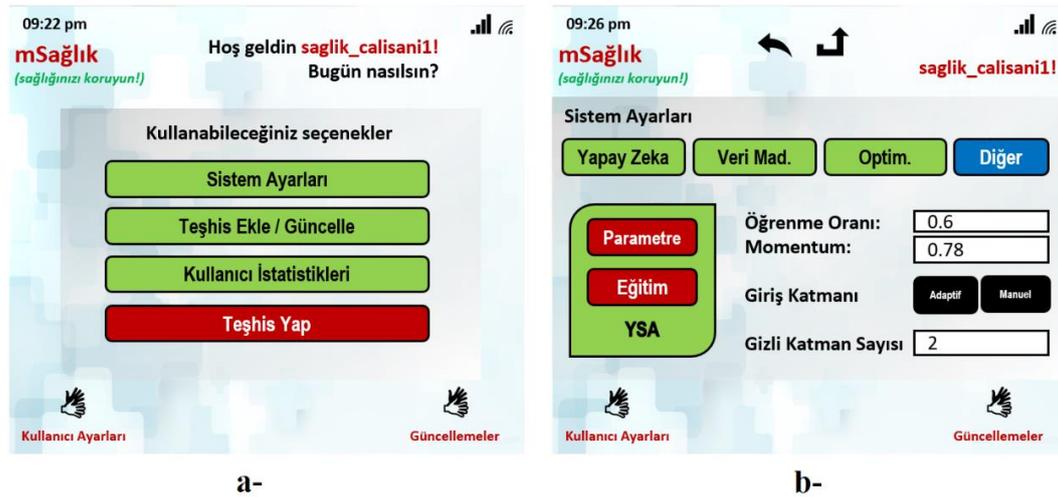

Figure 1. a- Healthcare user menu screen. b- mHealth healthcare user supportive technologies setting screen.

Healthcare user can set the parameters for Artificial Intelligence, Data Mining and similar supporting technologies by selecting the "system settings" option. On the same screen, there is also the "training" option, which enables the training process to be carried out by uploading new training data to the system and at this point to update the AWS cloud system.

Options such as 'user settings', 'updates' in mHealth interface are generally standard options in mobile applications. Apart from these, healthcare users can access the interface where they can add diagnosis / expert information to the system by selecting the 'add / update diagnosis' option on the main menu screen (Figure 2).

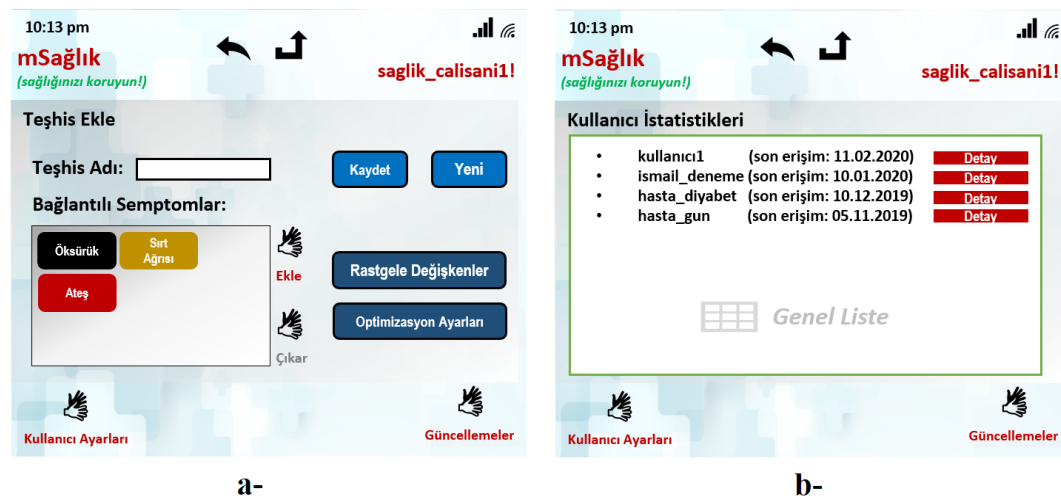

Figure 2. a-mHealthcare user 'add / update diagnosis' interface. b- mHealth healthcare user 'user statistics' interface.

As stated before, healthcare user can use the "diagnose" option on the main screen in order to test the system or even make diagnoses within the scope of their personal health processes. Explanations regarding this option, which is common with standard users, are given in the following paragraphs.

### 3.2.2. Standard user interfaces and usage

Standard users can log in to the mHealth system on the same screen, just like healthcare users. The menu screen presented to standard users after the login process is different from the screen presented to healthcare users, but it is supported with fewer options in order to ensure easy and fast use (Figure 4.5. A-). Apart from the menus on the screen in question, the standard options in mobic applications are also valid for standard users, but the most critical option that can be accessed from the main menu screen is the "diagnose" option. After selecting this option, the user is asked in which way he/she will switch to the diagnostic process (Figure 3).

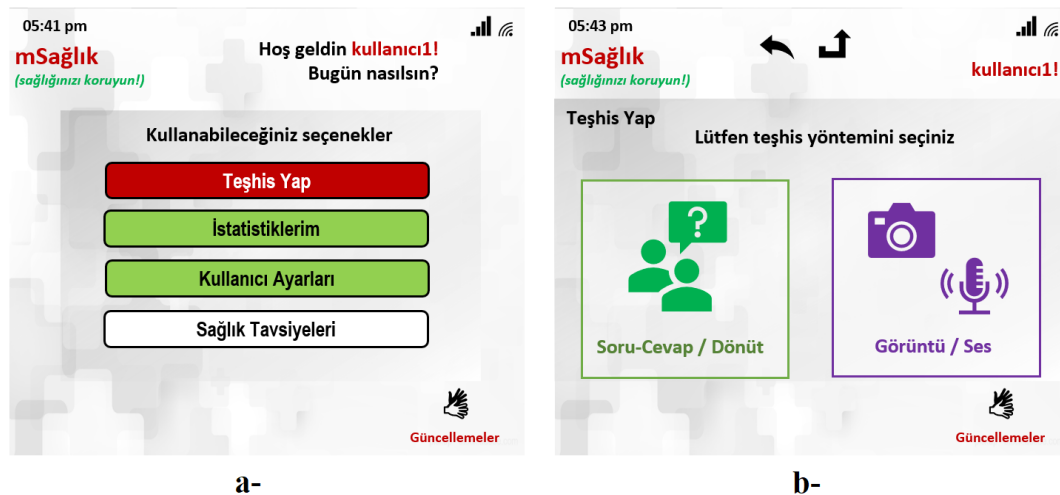

Figure 3. a-mHealth standard user menu screen. b- mHealth standard user 'diagnose' interface.

When the users select the "question-answer" option in the diagnostic interface, the mHealth system continues the diagnosis process by following the expert system process and asking the user for basic complaints (Figure 4. a-). If the user wants to report a complaint that is not included in the options, he can define it in the system environment.

By using the smart-keyboard feature, the system can display the defined complaints close to the complaint specified by the user. If the user wishes to perform image or sound-based diagnostics directly, he has to select the relevant option. After choosing this option, mHealth system uses the interaction strategy again and goes to receive images or sound from the user (Figure 4).

According to the diagnosis processes performed, the mHealth system can make appropriate diagnoses by running the classification - clustering - optimization processes from the cloud environment. If there is more than one possibility in these diagnoses, the relevant diagnoses can be expressed in the interface. If the diagnosis is above the level of the system and requires advanced examination and intervention, the mHealth system directs the user to visit a hospital for further examinations. Figure 5 shows sample screenshots of feedback and image-oriented diagnosis results.

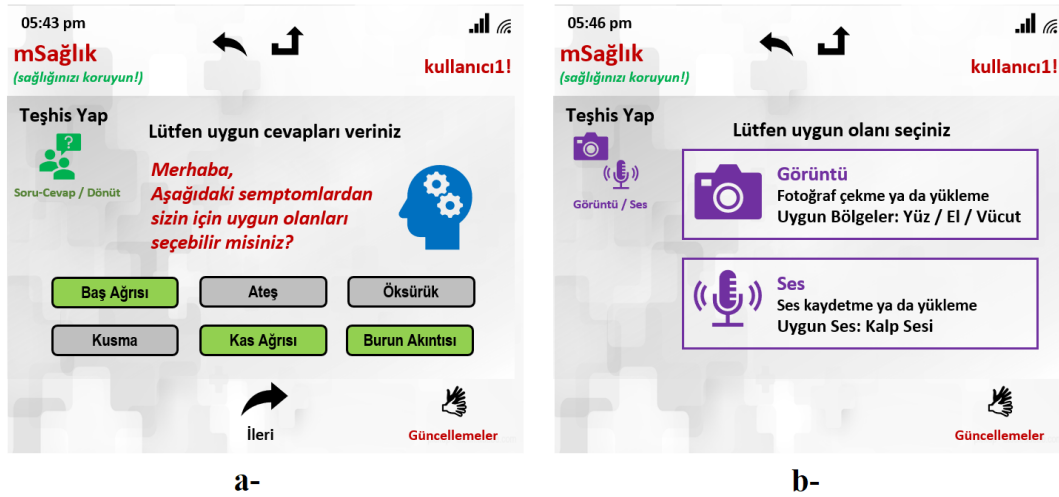

Figure 4. a-mHealth standard user feedback (question-answer) diagnosis. b- mHealth standard user image / sound-based diagnosis.

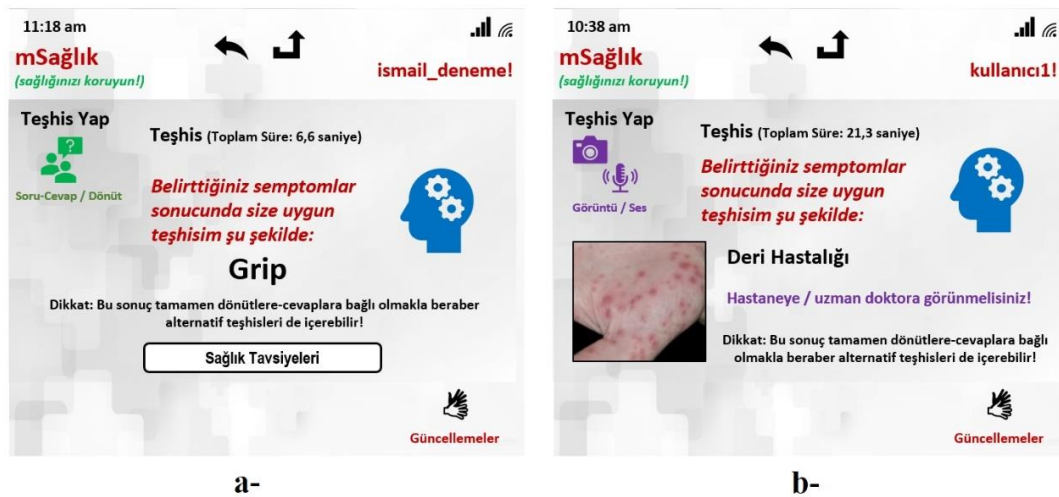

Figure 5. Examples of mHealth feedback and image-oriented diagnostic results.

With the help of the interfaces and options explained, the use of mHealth can be performed quickly and effectively within the scope of both users. The use of the aforementioned controls in the interfaces and the use of similar mechanisms - visuals are used to provide the best possible usage experience to all types of users (especially standard users who will diagnose).

## 4. EVALUATION

Here a general evaluation done over the mHealth system was reported accordingly. In this context, findings for the survey-based evaluation and diagnosis applications were explained generally.

### 4.1. Survey Based Evaluation

It is extremely important to use and evaluate the mHealth system from the eyes of healthcare staff. For this purpose, the Usage Survey, which includes a total of 10 expressions for the usage process, and the Process Survey consisting of 6 statements in total, were collected from a total of 10 people consisting of 4 specialist doctors and 6 health sector employees. The 5-point Likert Scale (1:

Strongly Disagree, 2: Disagree, 3: Undecided, 4: Agree, 5: Totally Agree) was used to collect the opinions. The average values obtained with the Likert Scale were taken into account when evaluating the system from the eyes of healthcare staff. The expressions in both surveys generally question the effectiveness and efficiency of the system in terms of health assistance and usage processes. Statements within the scope of the surveys and findings regarding the feedback received are presented in Table 3 and Table 4.

Table 3. Findings of the healthcare users Usage Survey.

| No | Statement | Feedback* | | | | | Mean |
|---|---|---|---|---|---|---|---|
| | | 1 | 2 | 3 | 4 | 5 | |
| 1 | "mHealth system is easy to use in general." | 0 | 0 | 0 | 2 | 8 | 4,8 |
| 2 | "mHealth system generally works fast." | 0 | 0 | 2 | 3 | 5 | 4,3 |
| 3 | "This system is suitable for personal health support use." | 0 | 0 | 0 | 3 | 7 | 4,7 |
| 4 | "Thanks to this system, individuals can fix simple health problems when necessary." | 0 | 0 | 1 | 3 | 6 | 4,5 |
| 5 | "I found the mHealth system difficult and boring to use." | 5 | 4 | 1 | 0 | 0 | 1,6 |
| 6 | "In the system, diagnoses for health problems are made consistently." | 0 | 0 | 1 | 2 | 7 | 4,6 |
| 7 | "This system can be developed for advanced examinations." | 0 | 0 | 1 | 3 | 6 | 4,5 |
| 8 | "This system is also suitable for use by healthcare professionals." | 0 | 0 | 1 | 4 | 5 | 4,4 |
| 9 | "Evaluations of mHealth system through images are effective." | 0 | 1 | 1 | 2 | 6 | 4,3 |
| 10 | "Evaluations of mHealth system based on user feedback are effective." | 0 | 0 | 1 | 3 | 6 | 4,5 |
| 11 | "Evaluations of mHealth system over sound data are effective." | 0 | 0 | 1 | 2 | 7 | 4,6 |

**\* Likert Scale: 1: Totally Disagree, 2: Disagree, 3: Undecided, 4: Agree, 5: Strongly Agree**
**\* Total 10 Users**

Table 4. Findings of the healthcare users General Process Survey.

| No | Statement | Feedback* | | | | | Mean |
|---|---|---|---|---|---|---|---|
| | | 1 | 2 | 3 | 4 | 5 | |
| 1 | "I had difficulty using this system." | 6 | 3 | 1 | 0 | 0 | 1,5 |
| 2 | "The management of the mHealth system was easy." | 0 | 0 | 0 | 3 | 7 | 4,7 |
| 3 | "The infrastructure adjustments of the mHealth system were easy." | 0 | 1 | 2 | 2 | 5 | 4,1 |
| 4 | "I was able to add any additional diagnoses and problems I wanted to the mHealth system." | 0 | 1 | 1 | 3 | 5 | 4,2 |
| 5 | "Internet communication speed of mHealth system was at an acceptable level." | 0 | 0 | 2 | 3 | 5 | 4,3 |
| 6 | "I found the overall performance and performance of this system bad." | 6 | 3 | 1 | 0 | 0 | 1,5 |

**\* Likert Scale: 1: Totally Disagree, 2: Disagree, 3: Undecided, 4: Agree, 5: Strongly Agree**
**\* Total 10 Users**

As can be seen from the relevant surveys, the general use of the mHealth system for healthcare users, the experienced process; Efficiency and productivity levels have been highly acceptable, especially in the context of detecting health problems and support. In these surveys, especially the determination of health problems of the system and their guidance, family management is very critical. The positive findings obtained for healthcare users who are in a position to evaluate the efficiency and productivity of the system are important as they are among the factors that prove the success of the system.

Apart from evaluating the mHealth system from the eyes of healthcare users, evaluations were also made for the use of 15 different users (subjects) selected. Accordingly, unlike the surveys used by healthcare users, some survey items were arranged for personal use. Relevant surveys were organized to give an idea about the use and process efficiency and performance of the mHealth system, and feedbacks were received within the scope of the Likert Scale. At this point, while the Personal Use Survey consists of 8 statements in total, the Personal Process Survey was supported with 5 statements. Findings obtained with the expressions of Personal Use Survey and General Process Survey Table 5 and Table 6.

Table 5. Findings of the Personal Use Survey.

| No | Statement | Feedback* | | | | | Mean |
|---|---|---|---|---|---|---|---|
| | | 1 | 2 | 3 | 4 | 5 | |
| 1 | "I was able to use the mHealth system without any problems." | 0 | 0 | 2 | 4 | 9 | 4,5 |
| 2 | "MHealth system generally works fast." | 0 | 1 | 2 | 5 | 7 | 4,2 |
| 3 | "I found the mHealth system difficult and boring to use." | 7 | 5 | 1 | 2 | 0 | 1,9 |
| 4 | "The health support I received with this system was effective." | 0 | 0 | 2 | 6 | 7 | 4,3 |
| 5 | "I found the interface of this system informative and easy to use." | 0 | 0 | 1 | 4 | 10 | 4,6 |
| 6 | "There may be different health problems in this system." | 0 | 0 | 1 | 3 | 11 | 4,7 |
| 7 | "I found the suggestions offered with this system effective and successful." | 0 | 1 | 2 | 5 | 7 | 4,2 |
| 8 | "I recommend mHealth system to my friends - to my environment." | 0 | 0 | 2 | 3 | 10 | 4,5 |

\* Likert Scale: 1: Totally Disagree, 2: Disagree, 3: Undecided, 4: Agree, 5: Strongly Agree
\* Total 15 Users

Table 6. Findings of the Personal General Process Survey.

| No | Statement | Feedback* | | | | | Mean |
|---|---|---|---|---|---|---|---|
| | | 1 | 2 | 3 | 4 | 5 | |
| 1 | "I had a hard time using this system." | 11 | 3 | 1 | 0 | 0 | 1,3 |
| 2 | "It was easy to express my problems in the mHealth system." | 0 | 0 | 1 | 4 | 10 | 4,6 |
| 3 | "User adjustments of mHealth system were easy." | 0 | 0 | 1 | 6 | 8 | 4,5 |
| 4 | "I found the overall performance and performance of this system bad." | 8 | 4 | 1 | 1 | 1 | 1,9 |
| 5 | "Internet communication speed of mHealth system was at an acceptable level." | 0 | 1 | 2 | 6 | 6 | 4,1 |

\* Likert Scale: 1: Totally Disagree, 2: Disagree, 3: Undecided, 4: Agree, 5: Strongly Agree
\* Total 15 Users

As can be understood from the findings obtained, standard users also find the general use of the mHealth system and the processes they experience effective and successful. At this point, the consistency of positive feedback given by normal users and healthcare users to statements with the same structure is among the points that confirm the performance of the system. The system is easy to use in general terms, fast, effective enough in terms of health support and working at an acceptable level in Internet communication processes were among the important factors with positive findings.

### 4.2. Diagnosis Evaluation

The next applications in the mHealth system evaluation processes have focused on diagnosis, especially in order to show the hit rate of Artificial Intelligence and other supporting technologies. Accordingly, diagnosis processes were carried out by taking into account the scenarios requiring advanced examination, user feedback (taking into account both certain and uncertain parameters), image data and even sound data, together with daily basic health problems. Findings related to different scenarios - 30 different trials performed for applications - diagnosis are given in Table 7 together with the details of each application.

Table 7. Findings regarding mHealth diagnosis.

| No | Problem-Application Detail* | Mean Success / Accuracy |
|----|------------------------------|--------------------------|
| 1 | Flu | 100% |
| 2 | Cold | 100% |
| 3 | Flu (+2 uncertain parameters) | 98% |
| 4 | Vitamin Deficiency | 100% |
| 5 | Osteoporosis | 97% |
| 6 | Goiter (with +3 uncertain parameters) | 95% |
| 7 | Cyst symptom (hospital / specialist doctor recommendation) | 88% |
| 8 | Normal heart sound | 94% |
| 9 | Normal heart sound (noisy) | 89% |
| 10 | Abnormal heart sound | 92% |
| 11 | Abnormal heart sound (loud) | 87% |
| 12 | The appearance of acne (hand) | 100% |
| 13 | Sweating image (face) | 98% |
| 14 | Measles image (body) (hospital / specialist doctor recommendation) | 95% |
| 15 | Healthy individual image (face) | 99% |
| 16 | Healthy individual image (hand) | 100% |
| \* In total 20 different running | | |
| \* Samsung Galaxy S10e mobile phone using Android 9.0 Pie operating system (6 GB dynamic memory, 128 GB hard disk memory) | | |

As it can be seen from the table, mHealth system can provide sufficient levels of successful feedbacks for different diagnostic scenarios - applications. Based on these findings, it is possible to state that the diagnostic performance criterion, which is critical for the mHealth system, has been met sufficiently.

### 5. CONCLUSIONS AND FUTURE WORK

Mobile health applications are among the important applications that have increased in popularity over time, especially as a result of our frequent use of mobile devices. Although we are in a period where entertainment and educational mobile applications are more popular, the need for a mobile application that increases personal productivity and supports us in an important issue such as our health among the rush of daily life seems extremely critical. From this point of view, it is possible to state that mHealth application offers a solution that meets the needs. Although there are similar applications in the literature, the fact that mHealth includes Expert System, ANN, intelligent optimization, signal - image processing and Data Mining technologies have caused it to contain different aspects from its similar ones. In order for the application to work effectively with all these technologies, cloud technology has been used, and the application, which aims to keep the user experience at a high level, basically based on the Expert System logic, has been supported with important Artificial Intelligence and data processing techniques - algorithms, making it more flexible and efficient.

The ANN technique included in mHealth has provided a successful infrastructure for updating expert knowledge and generating effective solutions to complex problems thanks to its flexible model structure. In this way, a cloud-connected Machine Learning approach has been provided to achieve a self-learning and developing health assistant system. The Machine Learning infrastructure, combined with the Data Mining infrastructure, has enabled the creation of an Artificial Intelligence background that can use all three of consultancy, non-consultancy and reinforcement learning. Processes such as directing the processes within the mHealth application, determining the most suitable processes that can be applied to change the current status of the user, or determining the most suitable among multiple diagnosis-solutions are also supported by intelligent optimization algorithms that can provide faster feedback. The Particle Swarm Optimization (PSO) and Ant Colony Optimization (CUR) were used for this purpose and increased the Artificial Intelligence capabilities of the application. Finally, signal-image processing processes, especially as a result of functions such as sensors and photo-taking features of mobile devices, have ensured that visual and audio data are also taken into the application environment at the point of personal health assessment. Today's Artificial Intelligence-based mobile applications include such data processing processes beyond their standard uses. Therefore, it has been an important advantage that mHealth has an infrastructure for signal-image processing capabilities.

Thanks to its knowledge and rule bases, mHealth can provide effective feedback especially against common diseases in daily life. By using various personal health data owned by the user, it is ensured that the personal health assistance processes include daily recommendations and directions. Although mHealth application's own internal functioning and interaction with the user is surrounded by optimization-based mechanisms that continue in the background, the mechanisms to offer solutions to more complex health problems and the evolution of expert knowledge on the system cloud side with the contribution of the user and the doctor-health staff allow the system to go a little further than simple health assistants. gives. Considering the findings obtained from the evaluation works carried out in order to understand whether all these advantages actually provide application performance, the following results can be mentioned:

**(1)** Application-oriented use and general process surveys have been effective tools to get opinions of both healthcare users and normal users regarding the application. In this context, healthcare users (especially doctors with the role of developing expert knowledge) who use the application think that mHealth is a tool that is easy to use and compatible with changing and developing conditions. **(2)** Again, doctors have positive feedback that the system is consistent in diagnoses made in various dimensions. **(3)** Feedback given by normal users to usage and general process surveys highlights the conclusion that mHealth is an effective and efficient tool. **(4)** Diagnosis performances of the system showed positive outcomes for health problems seen daily life.

There are also some future studies planned for the development and improvement of mHealth application. It is possible to explain the planned future works as follows:

• Although the application can use a number of signal-image processing solutions, it is planned to develop more capabilities in this direction in the future.

• In practice, alternative ANN architectural structures and even alternative hybrid structures to be created with ANN and different Artificial Intelligence techniques will be used and their contribution towards success will be evaluated.

• The expert knowledge infrastructure of mHealth will continue to be strengthened and the effectiveness and efficiency of the application will continue to be evaluated under different conditions (hospital, city, country, time management).

• Versions of mHealth compatible with different mobile operating systems such as iOS and Windows Mobile will also be developed.

• By working separately on mHealth's communication with the cloud, Machine Learning processes and other factors affecting the working process, it will be seen whether the working hours can be taken to a better level.

• The use and general process evaluations of mHealth at the scale of healthcare users and normal users will continue, according to the feedback; If needed, small and medium scale improvements will be made.

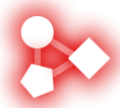

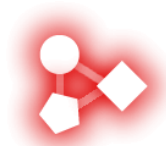